\documentclass[aps,showkeys,showpacs,onecolumn]{revtex4}
\usepackage{graphicx}
\usepackage{epstopdf}
\usepackage{psfrag}
\usepackage{amsmath,amssymb}
\begin{document}
\title{Muon-electron lepton-flavor violating transitions: shell-model calculations of 
transitions in $^{27}$Al }
\author{J.Kostensalo}
\affiliation{University of Jyvaskyla, Department of Physics, P.O. Box 35,
FI-40014, Finland}
\author{O.Civitarese}
\affiliation{Department of Physics,University of La Plata, and IFLP-CONICET, 
Argentina, 49 y 115. c.c.67 (1900), La Plata. Argentina}
\author{J.Suhonen}
\affiliation{University of Jyvaskyla, Department of Physics, P.O. Box 35,
FI-40014, Finland}

\begin{abstract}
In this paper we present the results of large-scale shell-model calculations of 
muon-to-electron 
lepton-flavor violating transitions for the case of the target nucleus $^{27}$Al. We
extend the previous shell-model calculations, done in the $sd$ model space, by including 
also the $p$ orbitals in order to see whether the negative-parity states produce any
significant effect in the conversion rate. 
The analysis of the results shows the dominance of coherent transitions mediated by 
isovector operators and going by the ground state of the target, with practically 
null influence of excited positive- or negative-parity states.
\end{abstract}
\maketitle
\section{Introduction}

The non-zero mass of the neutrino allows for a variety of lepton-flavor violating 
processes, forbidden in the standard model of particle physics. One of these many 
processes is $\mu \to e$ conversion \cite{Marciano1977,Marciano1977b,Vergados1986}, 
in which a bound $1S$ muon is captured by the nucleus and an electron is emitted with 
energy $E_{e}\approx m_{\mu}$. Other lepton-flavor violating processes include 
$\mu \to e\gamma$, $\mu \to e\bar{e}e$, $\tau \to \mu\gamma$, and $\tau \to e\gamma$. 
In minimal extensions of the standard model, in which massive neutrinos are included, 
the lepton-flavor violating processes would be suppressed by the ratio of the
masses of the neutrino and the weak boson,
$(m_{\nu}/m_W)^4 \sim 10^{-48}$--$10^{-50}$, and therefore be experimentally 
unobservable. The observation of such a conversion would therefore point to 
the existence of new massive particles, as was pointed out in \cite{Cirigliano2017}.

So far, only upper limits for the branching ratio 
\begin{equation}
R_{\mu e^-}= \frac{\Gamma (\mu^- \to e^-)}{\Gamma (\mu^- \to \nu_{\mu})}
\end{equation}
have been determined experimentally. The measured upper limits are at the 
90 \% confidence level $R_{\mu e^-}(\rm Au)<7\times10^{-13}$ for gold \cite{Bertl2006}, 
$R_{\mu e^-}(\rm Ti)<6.1\times10^{-13}$ \cite{Wintz1998} for titanium, and 
$R_{\mu e^-}(\rm Pb)<4.6\times10^{-11}$ for lead \cite{Honecker1996}. The Comet \cite{comet} 
and Mu2e \cite{mu2e} experiments aim at reaching single-event sensitivities 
$\sim 2.5 \times 10^{-17}$, corresponding to $< 6\times10^{-17}$ upper limit at 
90 \% confidence level. For muon-flavor violating processes the most restrictive 
limit is currently $R_{\mu \to e\gamma}<4.2\times10^{-13}$ \cite{Baldini2016}. 
For $\tau$-flavor violating decays the experimental upper limits are several 
magnitudes higher, with $R_{\tau \to e\gamma}<3.3\times 10^{-8}$ and 
$R_{\tau\to\mu\gamma}<4.4\times 10^{-8}$ \cite{Aubert2010}.

The nuclear matrix elements related to the $\mu^- \to e^-$ conversion have been 
previously studied in \cite{Siiskonen2000}. The nuclear-structure calculations were 
done using the nuclear shell model with the $sd$ shell as the valence space. The 
experimentally measurable coherent channel was found to be clearly dominant. The 
calculations also showed a clear vector-isoscalar dominance. 

In the present article we extend the shell-model valence space to include both the 
$p$ and $sd$ shells using a realistic effective interaction. In this way the potentially 
significant incoherent contributions with negative parity final states can be included. 
This also changes the occupation of the $p$-shell orbitals, which affects the important 
monopole part in the dominating coherent channel. The larger model space, however, 
increases the required computational burden significantly. For example the $m$-scheme 
dimensions for the $1/2^+$ states is 80 115 in the $sd$ shell but 61 578 146 in the 
$p-sd$ model space. Therefore, some limitations to the number of computed shell-model 
states are made. The coherent and incoherent contributions are calculated in order to 
find the ratio between the coherent and total $\mu^- \rightarrow \, e^-$ conversion 
rates. This ratio is needed, since experimentally only the coherent channel is 
measurable due to the lack of background events such as muon decay in orbit and 
radiative muon capture followed by $e^+e^-$ pair creation \cite{Dohmen1993}.

Since the formalism 
\cite{Kosmas1990,Kosmas1994,OC1994,Kosmas1996,Wild1984,Siiskonen1999,Kosmas1997b} is 
rather well known we shall restrict the presentation of it in Section \ref{formalism} 
to the minimum needed to allow for a comprehensible reading of the manuscript. 
In the following sections we shall introduce the definitions of the participant nuclear 
matrix elements as well as the basic ingredients of the shell-model procedure. The 
results of the calculations are presented and discussed in Section \ref{results} and 
our conclusions are drawn in Section \ref{conclusions}.

\section{Formalism}\label{formalism}

Various extensions to the standard model of particle physics allows for several possible 
mechanisms for $\nu^- \rightarrow \, e^-$ conversion, such as the exchange of virtual 
photons, a $W$ boson, or a neutral $Z$ boson \cite{Marciano1977,Marciano1977b,Vergados1986}, 
conversion by Higgs-particle exchange \cite{Kosmas1994,OC1994}, by supersymmetric 
(SUSY) particles \cite{Kosmas1997}, or by $R$-parity violating mechanisms \cite{Faessler2000}. 
The operators related to the hadronic verticies of the relevant Feynman diagrams, 
needed for the calculation of the $\mu^-\rightarrow e^-$ conversion mechanism mediated 
by interactions with nucleons, are of vector and axial-vector type. They can be written as
\begin{eqnarray}
{\bf{O}}_{\rm{vector}}&=&\sum_{k=p,n}C_{\rm vector}(k) e^{-iqr_k}
\nonumber \\
{\bf{O}}_{\rm{axial-vector}}&=&-\frac{1}{\sqrt{3}}\sum_{k=p,n}C_{\rm axial-vector}(k)(\sigma_k)e^{-iqr_k},
\end{eqnarray}
where the summations run over all nucleons and the couplings 
($C_{\rm vector}(k),C_{\rm axial-vector}(k)$) depend on the adopted mechanism for the  
$\mu^- \rightarrow e^-$ conversion \cite{Vergados1986}. Their values are given in
Table \ref{tab:coeff}. We use the same coupling constants as in the previous shell-model 
study \cite{Siiskonen2000}. Therefore, differences between the present and previous 
results are purely due to the differences in the nuclear-structure calculations,
and not blurred by adjustments made to the couplings.
\begin{table}[tbph]
\centering
\caption{Coefficients for each of the processes contributing to the $\mu^- \rightarrow e^-$ 
conversion in nuclei. The expressions of these coefficients are given in \cite{Siiskonen2000}.}
\begin{ruledtabular}
\begin{tabular}{ccccc}
Process  & $C_{\rm vector}({\rm neutrons}) $ & $C_{\rm vector}({\rm protons})$ & 
$C_{\rm axial-vector}({\rm neutrons})$ & $C_{\rm axial-vector}({\rm protons})$\\
\hline
$W$-exchange & 1.083 & 1.917& 0.017& -1.017  \\
SUSY $Z$-exchange &  3.230 & -0.230     &4.278 & -4.278 \\
photonic & 0.0& 1.0      & 0.0 & 0.0 \\
\end{tabular}
\end{ruledtabular}
\label{tab:coeff}
\end{table}

The square of the matrix elements of the vector and axial-vector operators entering 
the vertices where the leptonic and nucleonic currents exchange bosons \cite{Bilenky1987} 
are expressed in terms of the summation 
\begin{eqnarray}\label{nme}
M^2&=&S_{\rm V}+3\;S_{\rm A} \nonumber \\
S_{\rm V}&=&\frac{1}{(2J_{\rm{initial}}+1)}\sum_{\rm{final}}
\left(\frac{q_f}{m_{\mu}}\right)^2\sum_{\lambda} {\mid \langle f \mid \mid 
{\bf{T}}_{\lambda \lambda \gamma=0}\mid \mid i\rangle \mid}^2 \nonumber \\
S_{\rm A}&=&\frac{1}{(2J_{\rm{initial}}+1)}\sum_{\rm{final}}
\left(\frac{q_f}{m_{\mu}}\right)^2\sum_{\lambda \kappa} {\mid \langle f \mid \mid 
{\bf{T}}_{\lambda \kappa \gamma=1}\mid \mid i \rangle \mid}^2, 
\end{eqnarray}
where the initial and final states belong to the target nucleus and $m_{\mu}$ is the
muon rest mass. The tensor operators which result from the plane-wave expansion of
the factor $e^{-iqr_k}$ are of the form \cite{Bohr1969}
\begin{eqnarray}
{\bf {T}}_{\lambda\kappa\gamma,\mu}(qr)=j_{\kappa}(qr)
\left[{\imath}^{\kappa}Y_{\kappa}(\bf{\hat
r})\sigma_{\gamma}\right]_{\lambda\mu}.
\end{eqnarray}
In the particle representation they are written in  the form
\begin{equation}
{\bf
{T}}_{\lambda\kappa\gamma,\mu}(qr)=\frac{1}{\sqrt{2\lambda+1}} \sum_{f,i}
\langle f\mid\mid {\bf {T}}_{\lambda\kappa\gamma}(qr)\mid\mid
i\rangle \left[c^{\dagger}_f \overline{c}_i\right]_{\lambda\mu},
\end{equation}
where with $i$ and $f$ we denote the complete set of the
corresponding  quantum numbers needed to define a state in the
single-particle basis, that is $i(f)=(n,l,j,m)$. In standard
notation $n$ is the number of nodes, $l$ is the orbital angular
momentum and $(j,m)$ are the total angular momentum and its $z$
projection, respectively, of the single particle orbit. In the
previous equations the momentum transferred from the muon to the
nucleons is defined by $q_f=m_{\mu}-E_{\rm b}-E_{\rm{excitation}}$, that is
by the difference between the muon mass (105.6 MeV), the binding energy of 
the muon in the 1S orbit (0.47 MeV) and the excitation energy of each of the final
states of the nucleus which participate in the process, respectively.

The ratio of the experimentally measurable coherent channel and the total conversion rate is
\begin{equation}
\eta = \frac{\Gamma_{\rm coh}(\mu \to e^-)}{\Gamma_{\rm tot}(\mu \to e^-)} 
\approx \frac{M^2_{\rm coh}}{M^2_{\rm tot}},
\end{equation} 
where $M_{\rm coh}$ is the matrix element for the coherent transition and $M_{\rm tot}$ 
is the matrix element for the total conversion rate.

\section{Shell-model basis and interactions}

The wave functions and one-body transition densities needed for transitions between 
states in $^{27}\rm Al$ were computed assuming a $^4\rm He$ core and using the  
$0p$, $0d$, and $1s$ orbitals as the valence space for both protons and neutrons. 
The Hamiltonian adopted to perform the calculations, labelled \emph{psdmod}, was 
taken from Ref. \cite{psdmod}. It is a modified version of the interaction presented 
by Utsuno and Chiba in \cite{Utsuno2011}, which itself is a modification of the 
interaction PSDWBT of Warburton and Brown \cite{psdwbt}. In the parametrization given 
by the \emph{psdmod} Hamiltonian, the $p-sd$ shell gap has been increased by 1 MeV 
from the original interaction of Utsuno and Chiba. 
The calculations were performed using the shell-model 
code NuShellX@MSU \cite{nushellx} on a computer cluster using four 12-core Intel 
Xeon E5-2680 v3 @ 2.50GHz CPUs and took approximately 48 hours. For each spin and 
parity ten states were calculated. 

Including excitations over the $p-sd$ shell gap allows the description of the 
negative parity states, which can not be described using only the $sd$ shell. 
Therefore, the potentially significant spin-dipole contributions are included in 
the present study. 

\section{Results}\label{results}

\begin{figure}[tbhp]
\includegraphics*[width=8cm,height=8cm]{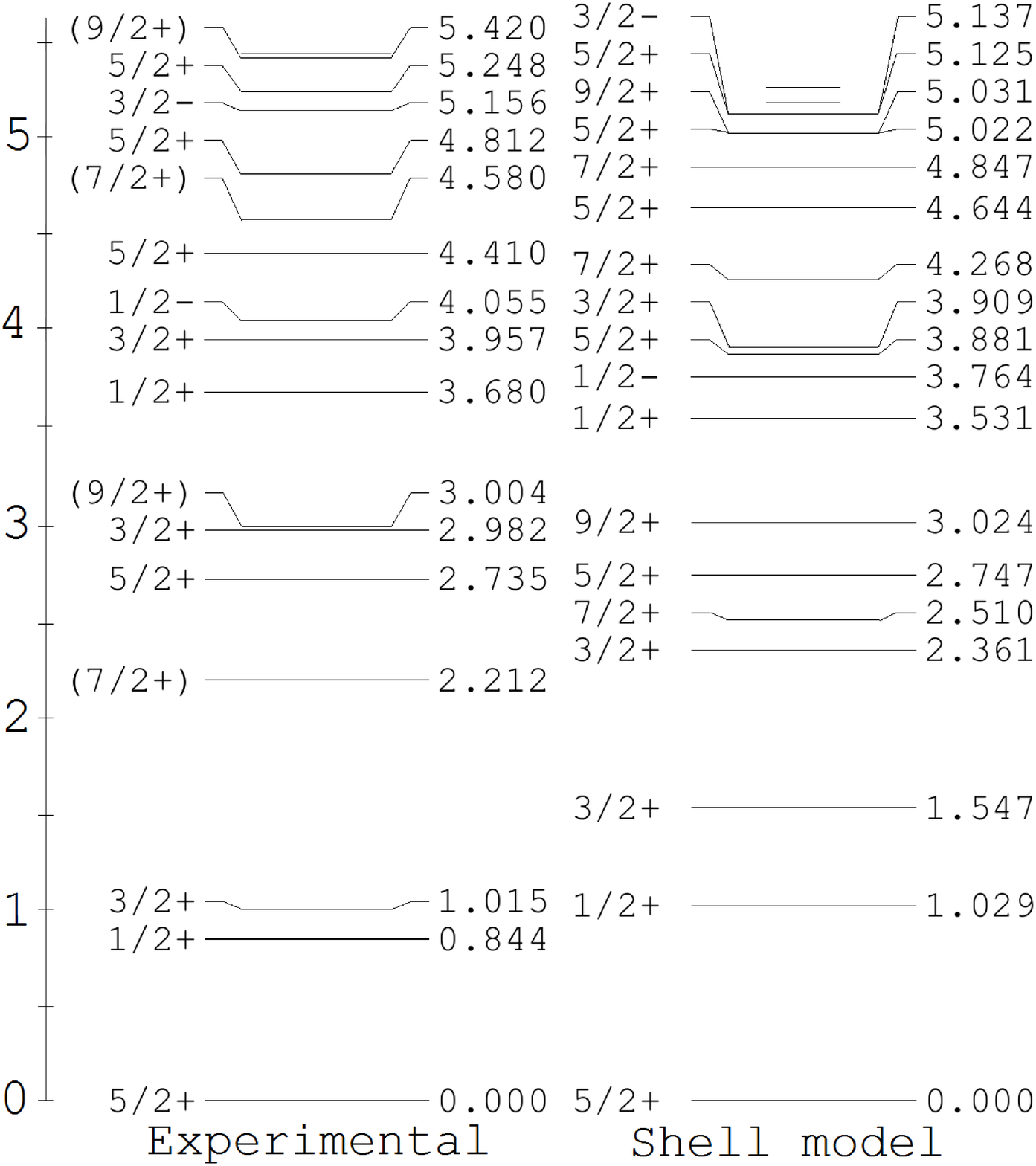}
\caption{Calculated (Shell-Model) and measured  (Experimental) energy levels of $^{27}$Al. The experimental energy 
spectrum is taken from Ref. \cite{nndc}. The energies are given in units of MeV.}
\label{fig:levels}
\end{figure}

The calculated spectrum of $^{27}$Al, with displayed states up to 6 MeV, is shown in 
Fig. \ref{fig:levels}, where the theoretical results are compared with the available 
experimental data. As seen from the figure, the energy gap between the $5/2^+$ ground 
state and the lowest excited states, which is of the order of 1 MeV is reasonably 
well reproduced, although the splitting between the first $3/2^+$ state
and the first $1/2^+$ state is larger in the calculations than in the data. 
For the rest of the spectrum both the sequence of the spin and parity values and 
the corresponding energies are reasonably well reproduced by the calculations.

As can be seen from Fig. \ref{fig:levels}, in addition to the good reproduction of 
the positive-parity states, the shell model is able to reproduce the energies of 
the lowest $1/2^-$ and $3/2^-$ states at 4055 keV and 5156 keV reasonably well, with 
calculated values of the energies at 3764 keV and 5137 keV, respectively. Not seen 
in the figure, the energy of the first $5/2^-$ state at 5438 keV is also well reproduced 
by the shell model, placing it at 5.433 keV. This is strong evidence that the size 
of the $p-sd$ shell gap is correct in the adopted Hamiltonian.

\begin{table}[tbhp]
\centering
\caption{Single-particle occupation factors $V_j^2$ for active orbitals, both for 
protons and neutrons.}\label{tsingle}
		\begin{ruledtabular}
\begin{tabular}{ccc}
Orbital   & $V_j^2$ (Protons)  & $V_j^2$ (Neutrons) \\
\hline
$1p_{3/2}$ &  3.983 & 3.983 \\
$1p_{1/2}$ & 1.971  & 1.973 \\
$2d_{5/2}$ & 4.237& 4.9629  \\
$2d_{3/2}$ &0.468 & 0.596  \\
$2s_{1/2}$ & 0.341 & 0.4837  \\
\end{tabular}
		\end{ruledtabular}
\end{table}

The shell-model occupancies for the single-particle states in the active orbitals 
are given in Table \ref{tsingle}. The $0s$ shell is taken to be fully occupied.
In the next table \ref{tablef} we show the results of the present calculations for 
the nuclear form factors $F_Z(q^2)$ and $F_N(q^2)$, calculated using the shell-model 
occupancies of Table \ref{tsingle}. The form factors are given by
 \begin{eqnarray}
 F_Z(q^2)&=& \frac{1}{Z}  \sum_j \sqrt{(2j+1)}\langle j||j_0(qr)||j \rangle {(V_j^p)}^2 
\nonumber \\
 F_N(q^2)&=& \frac{1}{N}  \sum_j \sqrt{(2j+1)}\langle j||j_0(qr)||j\rangle {(V_j^n)}^2, 
 \end{eqnarray}
where $p$ and $n$ denote proton and neutron states, while the coherent form factors 
$S_N=NF_N$ and $S_Z=ZF_Z$, where $N=14$ and $Z=13$ are the neutron and proton numbers 
in the case of $^{27}$Al. With respect to the previous shell-model results 
(see \cite{Siiskonen2000}), the present value of the coherent matrix element squared  
is smaller but still some 10 percent larger than the experimental value. 

\begin{table}[tbhp]
\centering
\caption{Ground-state values of the calculated and experimentally extracted matrix 
elements for $q/(\hbar c)=0.534$}
\begin{ruledtabular}
\begin{tabular}{cccc}
Quantity  & Experimental value&Shell model (Ref. \cite{Siiskonen2000}) & Shell model (present)\\
\hline
$F_N$ & 0.640 &  & 0.662 \\
$S_N$ &   &    & 9.264\\
$F_Z$ & 0.638&   0.677    & 0.667\\
$S_Z$ &     &       & 8.678 \\
$M^2_{\rm coherent} $&68.80&77.40&75.31 \\
\end{tabular}
\label{tablef}
\end{ruledtabular}
\end{table}

In Table \ref{process} we are listing the calculated values for the matrix elements 
of Eq.~(\ref{nme}), for each of the processes which may contribute to the muon-to-electron 
conversion. From the results given in the table it is seen that the calculated values 
are practically exhausted by the ground-state contributions to each process, and 
that the spin-independent vector channel is the dominant transition for all the 
processes considered. The analysis of the contributions to these matrix elements shows 
that they are given in almost equal parts by proton and neutron configurations, and 
that the main contribution which yields a value practically equal to the final one 
is given by the ground state of the target nucleus.

\begin{table}[tbhp]
\centering
\caption{Vector $(S_{\rm V})$ and axial ($S_{\rm A}$) matrix elements. In the
table we show the ground-state (g.s.) and total contributions to each quantity
as well as the relevant matrix element $M^2$.}
\begin{ruledtabular}
\begin{tabular}{cccccc}
Process  & $S_{\rm V}$(g.s.) & $S_{\rm V}$(total) & $S_{\rm A}$(g.s) & $S_{\rm A}$(total)
& $M^2=S_{\rm V}+3S_{\rm A}$\\
\hline
$W$-exchange & 664.9 & 665.1& 0.0& 0.371 &666.20\\
SUSY $Z$-exchange &  743.5 & 743.9   & 0.0 & 6.765 &764.20\\
photonic & 69.66& 69.68     & 0.0 & 0.0 &69.68\\
\end{tabular}
\end{ruledtabular}
\label{process}
\end{table}

\section{Conclusions}\label{conclusions}

The measurement of muon-to-electron conversion process in nuclei is a convenient tool 
to establish limits on the lepton-flavor violation. The nucleus $^{27}$Al is one of 
the nuclei of choice for performing the experiments, and due to it we have calculated 
the nuclear matrix elements for the operators entering the muon-to-electron conversion 
mediated by W-exchange, SUSY particles, Z exchange and photon exchange between nucleus 
and leptons with production of neutrinos. From our results, based on large-scale
shell-model calculations of the wave functions of the participant nuclear states in Al, 
we confirm the dominance of spin-independent, coherent ground-state transitions, 
with practically no contributions from the excited positive-parity or negative-parity
states. 
\section{Acknowledgements}
One of the authors (O.C) thanks with pleasure the hospitality of the Department of 
Physics of the University of Jyv\"askyl\"a. J.K. acknowledges the financial support 
from Jenny and Antti Wihuri Foundation. This work has been financed in part by the CONICET (Grant PIP-616) of Argentina and by the ANPCYT of Argentina. (O.C.) is a member of the CONICET of Argentina.

\bibliography{literature}

\end{document}